%% file: main.tex
\documentclass[copyright,creativecommons]{eptcs}
\providecommand{\event}{SYNT 2016} 
\usepackage{amsthm}
\usepackage{amssymb}
\usepackage{mathtools}
\usepackage{temporal}
\usepackage{enumerate} 
\usepackage{color}
\usepackage{tikz}
\usepackage{paralist}
\usepackage{fixme,manfnt,mparhack}
\usepackage[T1]{fontenc}
\usepackage{enumerate,enumitem}
\usepackage{xspace}
\usepackage{amssymb}
\usepackage{paralist}
\usepackage{wrapfig}
\usepackage{bbm}
\usepackage{cite}

\usepackage{macros}

\providecommand{\thisvolume}[1]{this volume of {\sl Electronic
  Proceedings in Theoretical Computer Science}. Open Publishing Association}

\definecolor{mygreen}{rgb}{0,0.6,0} 
\definecolor{mygray}{rgb}{0.9,0.9,0.9} 
\definecolor{mymauve}{rgb}{0.58,0,0.82}

\usepackage{listings}
\usepackage[finalnew]{trackchanges}
\addeditor{SJ}
\addeditor{RB}

\usepackage{xspace}

\input{preamble}

\title{The Reactive Synthesis Competition:\\ \syntcomp 2016 and Beyond}
\author{Swen Jacobs
\institute{Saarland University\\Saarbr\"ucken, Germany}
\and
Roderick Bloem
\institute{Graz University of Technology \\ Graz, Austria}
}
\def\titlerunning{\syntcomp: 2016 and Beyond}
\def\authorrunning{S. Jacobs and R. Bloem}
\begin{document}
\maketitle

\begin{abstract}
We report on the design of the third reactive synthesis competition (\syntcomp 2016), including a major extension of the competition to specifications in full linear temporal logic. We give a brief overview of the synthesis problem as considered in SYNTCOMP, and present the rules of the competition in 2016, as well as the ideas behind our design choices. Furthermore, we evaluate the recent changes to the competition based on the experiences with SYNTCOMP 2016. Finally, we give an outlook on further changes and extensions of the competition that are planned for the future.
\end{abstract}

\input{intro}

\input{problem}

\input{2016}

\input{eval}

\input{future}

\input{conclusions}

\bibliographystyle{eptcs}
\bibliography{synthesis}
\end{document}

%% file: preamble.tex
\newcommand{\myparagraph}[1]{\smallskip \noindent {\em #1.}}

\newcommand{\bad}{\textsf{Error}\xspace}

\newcommand{\syntcomp}{SYNTCOMP\xspace}

\newcommand{\simpleBDD}{Simple BDD Solver\xspace}

\newcommand{\acacia}{Acacia\xspace}

\newcommand{\slugs}{\texttt{slugs}\xspace}

\newcommand{\party}{\textsc{Party}-Elli\xspace}

\newcommand{\unbeast}{\textsc{Unbeast}\xspace}

\newcommand{\subby}{\kern-0.2em\leftarrow\kern-0.2em}

%% file: intro.tex
\section{Introduction}
\label{sec:intro}

The automatic synthesis of reactive systems from formal specifications has 
been one of the major challenges of computer science for more than 50 years, 
and a number of fundamental approaches to solve the problem have been 
proposed~\cite{Church62,BL69,Rabin72,PnueliR89}. For a long time, 
the impact of theoretical results on the practice of system design has been 
rather limited, due to the high worst-case complexity of synthesis from 
specifications in expressive temporal logics, and a lack of algorithms that 
solve the problem efficiently in the average case. Recently, there have been 
a number of new approaches that aim at more practical synthesis algorithms by 
either restricting the specification 
language~\cite{BloemJPPS12,MorgensternS11}, 
or by a smart exploration of the search 
space~\cite{Finkbeiner13,Ehlers12,fjr11,SohailS13,FinkbeinerJ12,FiliotJR13}. 
Moreover, there has been an 
increased interest in applications of reactive synthesis techniques, e.g., in 
robotics and cyber-physical systems, or for the synthesis of device drivers.
\cite{ChinchaliLTBM12,Kress-GazitFP09,ChenDSB12,LiuOTM13,RyzhykWKLRSV14} 
Despite this growing interest, there remains a divide between theoretical 
research and applications, due in some part to a missing infrastructure to 
compare synthesis tools, and a lack of incentive to build efficient and 
mature implementations (as noted by Ehlers~\cite{Ehlers11}).

In 2014, the authors and Ehlers founded the reactive synthesis competition 
(\syntcomp) in order to foster the research in scalable and user-friendly 
implementations of synthesis techniques.
The goals of \syntcomp are 
\begin{enumerate}[label=\roman*)]
\itemsep0pt
\item to make synthesis tools comparable by establishing a \emph{common 
benchmark format},
\item to facilitate the exchange of benchmarks in a \emph{public benchmark 
repository},
\item to establish a dedicated platform for a \emph{comprehensive and fair 
evaluation} of synthesis tools,
\item to encourage the implementation of synthesis tools that can be used as 
\emph{black-box solvers} in applications, and
\item to foster the \emph{efficient implementation} of synthesis algorithms 
by regularly providing new and challenging benchmark problems, and comparing 
the performance of tools on these.
\end{enumerate}
Since its inception, \syntcomp was held annually, and the first two 
iterations~\cite{SYNTCOMP14,SYNTCOMP15} were intentionally restricted to 
safety properties and a low-level specification format derived from the 
existing AIGER format~\cite{aiger,SYNTCOMP-format}, in order to have a low 
entry barrier for participants. We consider the competition to be a great 
success thus far: where before there were no two synthesis tools that used the 
same input language, there are now five tools from different research 
groups can be compared in a fair and 
meaningful way, based on a common specification language. As a part of the 
\syntcomp effort, we maintain a public benchmark library\footnote{Synthesis 
Competition Repository: \url{https://bitbucket.org/swenjacobs/syntcomp/}} 
which now consists of several thousand benchmark instances from a wide range 
of domains, and is steadily growing. 
Moreover, \syntcomp has triggered an increased interest in the development of 
efficient synthesis tools and specification languages that relate to the 
competition,
as witnessed by a growing number of publications on these 
topics~\cite{BloemKS14,BloemEKKLS16,BrenguierPRS14,BrenguierPRS15,Khalimov16}
\footnote{Moreover, the ideas from~\cite{WalkerR14} were also used in the 
version of \simpleBDD that competed in \syntcomp 2015.}, 
including tools and research groups that have not 
participated in the competition thus 
far~\cite{NarodytskaLBRW14,EenLNR15,Tentrup16,ChiangJ15,LeggNR16}.

\syntcomp 2016 introduces a major extension of the competition by dropping 
the restriction to low-level safety properties. To this end, we add a separate 
competition track for the evaluation of synthesis tools on specifications in 
a high-level input format for full linear temporal logic (LTL).
The specification format used for the new tracks is the \emph{temporal logic 
synthesis format} (TLSF), recently introduced by Jacobs and Klein~\cite{JacobsK16}. 

In this paper, we describe the design of \syntcomp 2016, with a focus on the 
extension to specifications in TLSF, and report on our plans for further 
extensions of the competition in the coming years. We describe the synthesis 
problem as considered in \syntcomp in Section~\ref{sec:problem}, followed by 
a presentation of the design and rules of \syntcomp 2016 in 
Section~\ref{sec:setup}. 
In Section~\ref{sec:eval}, we evaluate our recent changes to the competition, based on the experiences with \syntcomp $2016$.
Finally, in Section~\ref{sec:future} we 
give our thoughts on possible and probable further extensions of \syntcomp in 
the future, as a basis for discussion. 
Note that the benchmarks, participants and results of \syntcomp 2016 are presented in a sister paper~\cite{JacobsETAL16b}.

%% file: problem.tex
\section{Reactive Synthesis: A Brief Overview}
\label{sec:problem}

We briefly summarize the reactive synthesis problem as it is considered in \syntcomp, including approaches that have been developed to solve it. 

\paragraph{The Synthesis Problem.}
We consider the synthesis problem for reactive systems that can be 
represented as \emph{finite-state machines}. The specifications we consider come in 
two forms: either as temporal logic formulas, more specifically in \emph{linear-
time temporal logic} (LTL)~\cite{Pnueli77}, over the sets of inputs and outputs of the system, 
or as an AIGER circuit~\cite{aiger,SYNTCOMP-format} with a single output, with a set of controllable and a set of
uncontrollable inputs. 

For specifications in LTL, the \emph{realizability problem} is to decide 
whether there exists a finite-state machine that reads the inputs and 
produces outputs such that the specification is satisfied in all possible 
executions. For AIGER circuits, the \emph{realizability problem} is to decide 
whether there exists a controller circuit that reads the controllable inputs 
and the current state of the specification circuit, and produces the 
controllable inputs of the specification circuit such that the single output 
of the circuit is never raised. 

Given a realizable specification, the \emph{synthesis problem} is to find an 
implementation that satisfies the specification. 
The synthesis problems we consider are equivalent to finding a winning 
strategy in infinite two-player games whose structure and winning strategies are determined by the specification\cite{Thomas95}. For both kinds of 
specifications, solutions can be encoded into an AIGER circuit.

\paragraph{Important Fragments.}
There are several important fragments of LTL, differing in expressivity and 
in the complexity of the realizability and synthesis problems. For full LTL, 
the realizability and synthesis problems are 2EXPTIME-complete in the size of the specification 
formula~\cite{PnueliR89}. However, there exist a number of fragments for which the problems are decidable in EXPTIME~\cite{AlurT04,PnueliAMS98}, in particular the GR(1) fragment~\cite{BloemJPPS12}, which allows some restricted liveness 
properties in addition to simple safety properties.\footnote{More precisely, 
for GR(1) the size of the game arena is exponential in the size of the formula, and 
GR(1) games are solved in quadratic time in the size of the arena\cite{BloemJPPS12,Finkbeiner16}.}

\paragraph{Synthesis Algorithms.}
There are a number of existing algorithms to solve the synthesis problem, 
based on two fundamental approaches. The first approach, by B\"uchi and 
Landweber~\cite{BL69}, works by translation into a 
deterministic B\"uchi game, and solving it. The second approach, by 
Rabin~\cite{Rabin72}, works by 
translation into a tree automaton, and solving its emptyness 
problem.

In recent years, many algorithms for solving synthesis problems more 
efficiently have been proposed. We mention a few prominent approaches.
Bounded synthesis~\cite{Finkbeiner13} 
searches incrementally for solutions up to a certain size. An algorithm based 
on bounding liveness properties and a symbolic representation by antichains 
has been implemented in the synthesis tool \acacia~\cite{fjr11,acacia}.
Other algorithms try to exploit the structure of commonly occuring 
specifications, and propose incremental or compositional ways to solve the 
problem~\cite{SohailS13,fjr11,JobstmannB06}.

For safety properties, efficient algorithms can be implemented using BDDs and 
a fixpoint construction over the uncontrollable predecessors of the unsafe 
states. For GR(1), there is a similar algorithm, using a nested 
fixpoint construction~\cite{BloemJPPS12}. 
A more detailed introduction into approaches to solve safety games can be 
found in the report of \syntcomp 2014~\cite{SYNTCOMP14}.

%% file: 2016.tex
\section{\syntcomp 2016: Rules and Setup}
\label{sec:setup}
The basic idea of \syntcomp is that submitted 
tools are evaluated on a previously unknown selection of benchmarks from the publicly available library, without user 
intervention. Tools are then ranked with respect to the number of problem instances that can correctly be solved within a given timeout. The competition is separated into tracks that correspond to the fragments of LTL mentioned in Section~\ref{sec:problem}.

In the following, we first give an overview of the rules that are common to 
all tracks, and then go into some details for the separate tracks, with a 
focus on the changes made this year.

\paragraph{Tracks.} 
The competition is divided into two main \emph{tracks}, 
distinguished by the specification format: safety specifications in AIGER 
format, and full LTL specifications in TLSF.\footnote{A third track with GR(1) specifications in 
TLSF was planned for \syntcomp $2016$, but was not executed due to various reasons that we explain in Section~\ref{sec:GR1Track}.} 
In both tracks, realizability is defined with respect to Mealy 
semantics, i.e., the outputs of an implementation are allowed to depend on the 
inputs without any delay. 

In each track there are \emph{subtracks} for two different tasks: \emph{realizability checking} and \emph{synthesis}. While in \emph{realizability checking} the tools 
only need to return one bit of information, in \emph{synthesis} they need to 
return a provably correct solution. While the main goal of the competition is to compare algorithms for synthesis, we included subtracks for realizability checking to have a low entry barrier for participants. This decision is justified by our experience: in each of the competitions in $2014$, $2015$ and $2016$, there were $2$ tools that only supported realizability checking, but not synthesis.

Furthermore, for each (sub)track we separate the analysis of results into the two execution modes \emph{sequential} and \emph{parallel}. In \emph{sequential} mode, tools 
can only use one core of the CPU, and in \emph{parallel} mode they can 
use multiple cores in parallel. The idea of this separation is to have parallelization of algorithms as an explicit dimension of the competition. A side-effect in all iterations of \syntcomp thus far was that in sequential mode we only compared single algorithms, and portfolio implementations only appeared in the parallel mode (although the rules do not forbid to run portfolios in sequential mode). 

\paragraph{Entrants.} 
As in previous years, we ask participants to hand in their tools as source 
code, licensed for research purposes, accompanied by installation instructions and a short description of the 
system and the synthesis approach and optimizations it implements.

Each author can submit up to three different \emph{tool configurations} per 
subtrack. 
Our experience from previous iterations suggests that this limit is a good 
compromise that allows some flexibility for the tool creators, while 
avoiding the flooding of the competition with too many 
configurations of the same tool.

The organizers commit to making reasonable efforts
to install each tool, but reserve the right to reject entrants
where installation problems cannot be resolved. This was not the case
for any of previous iterations of the competition. 
In case of crashes or obviously wrong results we allow submission of 
bugfixes, if time permits. This possibility has been used in all previous iterations, including \syntcomp 2016.

We encourage participants to 
visit the SYNT workshop and the CAV conference for the presentation of the 
\syntcomp results, but this is not a requirement for participation. The 
organizers reserve the right to submit their own tools, and do so regularly.

\paragraph{Timeout.} In sequential execution mode, the timeout for each problem is $3600$s of CPU time. In the parallel mode, the timeout is $3600$s of wall time. 

\paragraph{Output Format.}
For the realizability checking tracks, tools should output either ``\texttt{REALIZABLE}'' or ``\texttt{UNREALIZABLE}'' on \texttt{stdout}.

For the synthesis tracks, tools should either output ``\texttt{UNREALIZABLE}'', or a circuit in AIGER format that satisfies the specification. In the safety-track, the specification has to be included in the solution, while in the LTL-track, the solution is only the synthesized strategy. 

\paragraph{Correctness of Solutions.}
Correctness in realizability subtracks is determined either by existing 
information about the realizability of the benchmark (possibly stored in the 
\texttt{STATUS} field of the specification~\cite{SYNTCOMP15}), 
or by a majority vote of all 
participating solvers if such information is not available. In the latter 
case, the execution platform for the experiments generates a notification 
that a previously unsolved problem has been solved, and the organizers 
inspect the problem to avoid errors in the evaluation. Correctness in the 
synthesis subtracks is determined by verification of the produced solution 
within a separate time limit of $3600$s (for details see below).

\paragraph{Ranking.}
Competition entrants are ranked with respect to the number of problems 
that can be answered with a correct solution within the given timeout. 
Timeouts (either in solving the problem or verifying the solution) are not counted, and 
wrong results are punished by subtracting $4$ points. Since all of 
the benchmarks are publicly available before the competition, such a punishment was not necessary in previous years. In \syntcomp 2016 however, we had one participant that returned a number of wrong results in the new TLSF-based synthesis track, and that could not be fixed in time for the competition (see the \syntcomp 2016 report~\cite{JacobsETAL16b}).
This ranking scheme for wrong solutions is also used in other competitions, but we have to agree with Cabodi et al.~\cite{cabodi2016hwmcc} that it is undesirable to possibly have a competition winner that produces a positive number of wrong results. Therefore, in future competitions we will disqualify tools that produce wrong solutions. If possible, we will allow the tool authors to supply a fixed version after the competition, and evaluate it hors concours, as we did this year. 

\paragraph{Quality Metrics.}
The goal of synthesis is to obtain implementations that are not only correct, 
but also efficient. Therefore, in previous iterations of \syntcomp we also 
considered additional \emph{quality rankings}, where correct solutions are 
additionally weighted based on their size. Since the rankings used in 
previous years gave unsatisfactory results, we do not have an official 
quality ranking this year. However, we still analyzed solutions with 
respect to their size and present our findings in the 
\syntcomp 2016 report~\cite{JacobsETAL16b}.

We plan to bring quality rankings back in future iterations of 
the competition, based on our experience from \syntcomp 2016 and our thoughts 
presented in Section~\ref{sec:future-quality}.

\paragraph{Competition Setup.}
Like in previous years, \syntcomp 2016 is organized on the EDACC platform~\cite{BalintDGGKR11}. The competition runs on a set of machines at Saarland University, each with a single Intel XEON processor (E3-1271 v3, quad-core, $3.6$GHz) and $32$ GB RAM (PC$1600$, ECC), running a GNU/Linux system. Each node has a local $480$GB SSD that can store temporary files. To ensure a  high comparability and reproducability of our results, a complete machine will be reserved for each job, i.e., one synthesis tool (configuration) running one benchmark. Since all nodes are identical and no other tasks will run in parallel, no other limits than the timeout will be set. 

\paragraph{Benchmark Selection.}
A subset of all available benchmarks was selected for the competition. 
Like in the previous year~\cite{SYNTCOMP15}, benchmarks are separated 
into categories, and we 
selected a subset from each category such that the different categories have approximately equal weight 
in the competition, and that the competition benchmarks represent a good 
distribution across different difficulties for each category. 

\subsection{Specific Rules for AIGER safety track}

\paragraph{Specifications.}
Specifications are given in the Extended AIGER Format for Synthesis~\cite{SYNTCOMP-format,SYNTCOMP14}, modeling a single safety property.

\paragraph{Output and Correctness.}
In the synthesis category, tools must produce solutions in AIGER format that include the specification circuit and abide by additional syntactic restrictions~\cite{SYNTCOMP14}. 
These are model checked with existing safety model checkers. 

Since model checking turned out to be a significant challenge for some problem 
instances in previous years, we introduce another extension in \syntcomp 2016. 
As an alternative to full model checking, tools can output, in addition to 
their solution, a \emph{winning region} of the system as a witness for 
correctness. If a winning region is 
supplied by the tool, we first try to verify correctness of the solution based on the invariant, and fall back to full model checking if the check is inconclusive.

\paragraph{Legacy Tools.}
For comparison, we run some of the entrants of \syntcomp 2014 and 
\syntcomp 2015 in the safety track. This allows us to highlight 
the progress of tools over the course of the last two years.

\subsection{Specific Rules for LTL Track}

\paragraph{Specifications.}
In the LTL track, specifications are given in basic TLSF format.
For tools that do not support TLSF directly, the organizers supply a number 
of translators to different existing formats in the SyFCo tool~\cite{SyFCo} 
(which will be installed on the competition machines). Specifications are interpreted according to standard LTL semantics. 

\paragraph{Output and Correctness.}
In the synthesis subtrack, tools must produce solutions in AIGER format if 
the specification is realizable. As a syntactical restriction, the sets of 
inputs and outputs of the TLSF file must be identical to the sets of inputs 
and outputs of the AIGER solution. Additionally, solutions are model 
checked with existing LTL model checking tools.

\paragraph{Legacy Tools.}
For comparison, we run the legacy synthesis tool \unbeast, non-competitive, in 
the LTL track. 
To this end, we convert the TLSF 
specification to the native input format of \unbeast, and used a 
wrapper script to make inputs and outputs conform to the standard format. 
Since this would be a significant amount of work for the synthesis subtracks, 
we use \unbeast only in the realizability subtrack.

%% file: eval.tex
\section{\syntcomp 2016: Evaluation of Changes}
\label{sec:eval}

We consider \syntcomp 2016, including the changes and extensions to the competition, as another successful step towards the overall goals of \syntcomp, as defined in Section~\ref{sec:intro}. In this section, we give a brief overview of our experience with \syntcomp 2016, separated into the continuation of the track based on safety specifications in AIGER format, and the introduction of the new track, based on full LTL specifications in TLSF.

\subsection{Existing Track: Safety Specifications}
One part of \syntcomp 2016 was the track with safety specifications in AIGER format. This track was already a part of \syntcomp 2014 and 2015~\cite{SYNTCOMP14,SYNTCOMP15}, with essentially the same setup as this year. More specifically, in \syntcomp 2016 the realizability subtrack ran with exactly the same rules as last year, and there were two changes to the rules in the synthesis subtrack: i) tools can now also supply a winning region in addition to their solution, in order to facilitate verification, and ii) the ranking is only based on the number of (provably) correct solutions, but not on their quality.

Regarding the number of participants and the progress over previous years, we think that it was well worth continuing the existing track, and that this will remain true for the foreseeable future: the number of participants increased to $6$ (from $4$ in \syntcomp 2015), and several tools performed significantly better than the best configurations from last year, that were run hors concours for comparison.

The possibility of supplying additional witness information to facilitate verification was used by $2$ out of $4$ tools that support synthesis of controllers, and proved to be very useful: out of the almost $1000$ solutions that were provided by the tools that used this option, only for $3$ the verification was inconclusive. In comparison, out of approximately $400$ solutions that came without this additional information, $14$ could not be verified.

Regarding the ranking, we think that the number of \emph{provably} correct solutions is a good basic measure that will also be used in the future. Additionally, we also plan to bring back a quality ranking (see Section~\ref{sec:future-quality}).

\subsection{New Track: Full LTL Specifications.}
The other main part of \syntcomp $2016$ was a completely new competition track that is based on specifications in full LTL. Specifications are given in the new \emph{temporal logic synthesis format} (TLSF)~\cite{JacobsK16}, that was designed to be clear and human-readable, to facilitate the development of complex, structured specifications, and can be automatically translated to a basic format that is easy to parse by synthesis tools.

We consider this new track a big success: the first iteration had $3$ participating tools, all of which supported not only realizability checking, but also synthesis. Additionally, we were able to run the legacy tool \unbeast on the competition benchmarks for comparison in the realizability checking subtrack. 

The newly designed format proved to be very well-suited for the competition, as witnessed by the following points:
\begin{itemize}
\item Specifications are easy to write, due to the structure of the format and the tool support. In particular, most of the benchmark instances used in the competition stem from explicitly \emph{parameterized} benchmarks that can automatically be instantiated for different valuations of their parameters.
\item The new specification format is also easy to integrate with existing synthesis tools. In \syntcomp 2016, versions of pre-existing synthesis tools \acacia and \party participated officially, and the legacy tool \unbeast was run hors concours. All of these used the \emph{synthesis format conversion tool} (SyFCo)~\cite{SyFCo}, supplied by the organizers and F. Klein, to automatically translate specifications to their native input formats. 
\end{itemize}

In the synthesis subtrack, verification of solutions by existing model checkers worked reasonably well. That is, for $2$ out of $3$ participants all of the solutions could be verified, while for the other one there were $20$ out of $153$ solutions where model checking was inconclusive. To facilitate verification, we will consider the possibility to include additional witness information with the solution. For more details, see Section~\ref{sec:witnesses}. 

%% file: future.tex
\section{The Future: Ideas for \syntcomp 2017 and Beyond}
\label{sec:future}

We discuss ideas for changes and extensions in upcoming iterations of \syntcomp. These are separated into six topics: a track for GR(1) specifications (Section~\ref{sec:GR1Track}), synthesis of compositional systems (Section~\ref{sec:compositional}), synthesis challenges (Section~\ref{sec:challenges}), quality ranking (Section~\ref{sec:future-quality}), witnesses for correctness (Section~\ref{sec:witnesses}), and the technical setup (Section~\ref{sec:future-setup}).

\subsection{A Track for GR(1) Specifications}
\label{sec:GR1Track}

As mentioned in Section~\ref{sec:problem}, there are a number of fragments of LTL that allow for more efficient synthesis procedures. The fragment that has found most application in practice is GR(1). We plan to add a separate track with specifications in GR(1) in \syntcomp 2017. Basically, the idea is to treat GR(1) as a fragment of LTL and use TLSF as the input format, while accounting for the non-standard semantics that is used with respect to safety assumptions in GR(1) (cp. Bloem et al.~\cite{BloemJPPS12}).

We already had plans for a GR(1) track in \syntcomp 2016, but these did not come to fruition due to several reasons. The main reason is that we were not able to collect a sufficiently large benchmark set for a meaningful comparison in time for the competition. This was in part due to a number of difficulties regarding the non-standard semantics, influencing what is deemed a correct solution, and how correctness can be checked. Additionally, there is an issue with the basic version of the TLSF format not being very well-suited for some important GR(1) benchmarks.

While not in time for \syntcomp 2016, we believe that we have now fixed the semantics issues, and a translation from formulas in the GR(1) semantics to formulas in the standard LTL semantics has been added to SyFCo, enabling automatic verification with LTL model checkers. Additionally, we were able to conduct some preliminary experiments that allow us to prepare better for the introduction of the track in 2017. In particular, we observed a much greater difficulty of verifying the solutions than for the existing tracks. With existing model checkers (that are tailored towards either safety or LTL specifications, but not towards GR(1)), model checking the results often takes much longer than producing them, making verification of all solutions infeasible. This means that for the GR(1) track we will either have to lift the requirement that solutions have to be provably correct, or will need additional witness information (see Section~\ref{sec:witnesses}) to make verification possible.

Regarding the specification format, the problem is that some of the specifications become very large when translated to the basic version of TLSF, which is currently the standard input formats for LTL synthesis tools in \syntcomp. The native input formats of GR(1) synthesis tools such as \slugs~\cite{EhlersR16} avoid the explosion of specification files by allowing sharing of sub-expressions in formulas. While this is already possible in full LTL, it is not allowed in basic LTL. To have a better support for GR(1), we plan a minor extension of TLSF with a format that is between the full and the basic formats in expressivity, and allows such sharing of subexpressions (and possibly some additional features).

\subsection{Compositional Specifications and Systems}
\label{sec:compositional}
We plan to extend the specification format and the competition to systems that consist of multiple components. Some possible extensions of the specification format are also discussed in 
the format description~\cite{JacobsK16}. Here, we focus on extensions 
of the competition, and what this means for the synthesis problems that need 
to be solved.

\paragraph{Compositional Specifications.}
Systems that need to be synthesized often consist of multiple components. These components 
can either be synthesized separately if specifications are completely local, 
or need to be synthesized such that the composed system additionally 
satisfies a global specification.
The latter case is interesting for \syntcomp, and is currently not supported 
by the specification format. 

\paragraph{Partial Implementations.}
When considering composed systems, a natural case of the synthesis problem is 
the synthesis of components for a system that is already partially 
implemented, i.e., where some components have a fixed implementation.

In some sense, this problem is already considered in the AIGER tracks of 
\syntcomp, as an AIGER file can contain both an implementation of a 
component, and a monitor automaton that raises an error output if the safety 
specification is violated. However, in an AIGER file there is no clear structural distinction between the two.

To give structural support for component-based systems, we consider an extension of TLSF that allows the specification of components that have a fixed implementation. Such an implementation of a component could 
for example be given as an AIGER circuit. The resulting format will  generalize both the existing TLSF 
format and the existing AIGER format, as explained in the following.

\paragraph{Integration of both formats into one.}
If the supported format includes both compositional TLSF specifications and 
partial implementations as AIGER circuits, then the resulting format 
generalizes both the existing TLSF format (obviously), and the existing AIGER 
format: a given specification in AIGER format can simply be added as a 
component with a fixed implementation and a single output \bad, where 
controllable inputs are assigned as outputs to the system to be synthesized, 
and the specification of the system is simply $$\always \neg \bad.$$


\paragraph{Imperfect Information:}
Finally, compositional specifications lead to the synthesis problem under 
partial information, i.e., the components need to decide on their behavior 
without knowing all inputs or the full internal state of the other 
components. As Pnueli and Rosner have shown~\cite{PnueliR90}, the synthesis 
problem is undecidable under partial information, even for safety 
specifications. However, there have been a number of approaches to solve 
instances of the 
problem~\cite{MadhusudanT01,MadhusudanT02,Finkbeiner13,FinkbeinerJ12,BloemCJK15}, 
and it would be interesting to include it into 
the competition at some point.

\subsection{Synthesis Challenges}
\label{sec:challenges}

In its third year, \syntcomp is still in the process of natural growth, and is only establishing itself as a regular institution in the synthesis community. In some related research fields, competitions have been around for a long time, and there have been some unintentional adverse effects on the development of tools. On the one hand, a competition gives additional incentive for the development of efficient push-button tools, and positive effects of competitions on the quality and efficiency of tools have been observed~\cite{JarvisaloBRS12,SutcliffeS06,BarrettDMOS13,cabodi2016hwmcc}. On the other hand, the specific design and rules of a competition may also discourage research on certain aspects of a problem, if they are \emph{not} part of the competition. A long-running competition may also produce a number of very efficient and mature tools that discourage newcomers from entering the field. 

Thus, as organizers of \syntcomp we have to admit to a responsibility for the research directions that we encourage or discourage by the design and the effects of the competition. 
One way to deal with the problems mentioned above would be flexible \emph{synthesis challenges} that change from year to year (or every few years), and might be decided on by the community. Some of the tasks mentioned in Sections~\ref{sec:compositional}, \ref{sec:future-quality} and \ref{sec:witnesses}, could be offered as challenges for a limited time. 

Another option is to provide potential participants with baseline solvers that already integrate the commonly accepted optimizations, such that the participants can focus on additional smart solutions, and don't have to implement all the basic features themselves. This approach could be enforced in a special track, where participants must start from this common baseline, and are only allowed to make limited changes to the implementation that is supplied. An example of such an approach are the ``Hack Tracks'' of the SAT competition~\cite{JarvisaloBRS12}, where participants start from a given SAT solver in source code, and the difference between the baseline and their own implementation is limited to $1000$ (non-space) characters.

\subsection{Quality Ranking/Quantitative Aspects}
\label{sec:future-quality}

As mentioned before, in synthesis we usually not only care about correctness 
of our implementations, but also about quantitative properties of the 
synthesized artifact, like its size, its reaction time to certain events, or 
possibly other aspects like energy efficiency.

\paragraph{Experience in Previous Competitions.}
In \syntcomp 2014 and 2015, we used different quality rankings based on the 
number of AND-gates in the solution, by comparing either against the size of 
other solutions in the given competition, or against the size of a reference 
solution. 
A comparison against a value that is not fixed before the competition means 
that the results (including the relative ranking of tools) may change when we 
add a tool. This is undesirable in general, and in particular if we 
want to use the results of the competition to evaluate a tool that did not 
participate. Therefore, using a reference solution is in general preferable. 
However, reference solutions are not always available, since one of our goals 
is to have \emph{new} and \emph{challenging} benchmarks in every iteration of 
the competition. Therefore, in \syntcomp 2015 we adopted a mixed solution, 
which uses the size of a reference solution if available, and the size of the 
smallest solution in the current run otherwise. Evaluation of 
non-participating tools then requires to take this distinction into account.

However, even with this mixed approach, the results of the quality ranking 
were somewhat unsatisfactory, since the size of solutions effectively only 
played a small role, and was dominated by the number of problems that could 
be solved. This was due to two main reasons. First, points for the 
size of solutions were given according to a $log_{10}$-scale, i.e., for 
solution $A$ to get one additional point compared to solution $B$, $A$ had to 
be $10$ times smaller than $B$. Probably a $log_2$-scale would be better if 
we want to emphasize the need for small implementations. Second, we compared 
the size of full solutions, which in the AIGER format includes the 
specification circuit. Since we have many problems with a very large specification, 
the size of the solution is often dominated by the size of the specification, 
and the size of the synthesized code does not make much of a difference. An 
option to repair this would be to compare the size of the synthesized code 
instead of the size of the full solution. 

In \syntcomp 2016, we experimented with these options, and some information can be found in the competition report~\cite{JacobsETAL16b}. In particular, it seems that comparing only the size of the actual implementation (or controller) gives significantly more meaningful results than comparing the sizes of the complete system that includes the specification circuit, and this will probably be the base of our quality ranking for \syntcomp 2017. On the other hand, the question how much a smaller solution should be rewarded, i.e., whether to use the $log_{10}$-scale, the $log_2$-scale, or some other measure, is a design decision that can not be answered objectively.

\paragraph{Quality of Solutions for LTL specifications.}
With our extension of \syntcomp to specifications in LTL, one question is 
whether the same ranking scheme is also suitable and fair for the new track. 
In the AIGER-based track, input and output are both symbolically encoded, and 
we can reasonably expect tools to optimize with respect to this encoding. In 
the LTL-based tracks, the input is not symbolically encoded, and the output 
encoding will in many cases be an additional step to conform to the 
competition format. Therefore, the answer to the question of fairness and 
suitability is not obvious. On the other hand, one can argue that almost all 
solutions that are efficient by some different measure can also be encoded 
into a small symbolic AIGER representation.  

In \syntcomp 2016 we 
experimented with the existing quality measures also for the TLSF-based track, 
and discussed the issue with the community at the SYNT workshop. In addition to the number of AND-gates, we also made experiments with the number of latches in the solution (since now solutions are in general not memoryless). Our results suggest that a ranking based on AND-gates will be similar to one that is based on latches: a solution with a high number of latches will almost always have a correspondingly high number of AND-gates. Since the number of latches is usually negligible compared to the number of AND-gates, one possibility is to not consider latches at all and use the same ranking as described above. Another option is to rank solutions according the combined size of AND-gates and latches in gate equivalents.

\paragraph{Quality Measures Beyond (Circuit) Size.}
There are many other natural quality measures for reactive systems. These include:
\begin{itemize}[itemsep=0mm]
\item the size of the reachable state space (either of the synthesized strategy, or of the solution that includes the specification circuit),
\item the reaction time (or delay) to certain actions/inputs of the environment, and
\item more complex measures that assign a cost to certain actions of the system, e.g. a measure for \emph{energy-efficiency} or \emph{power consumption}.
\end{itemize}

Specialized synthesis approaches that optimize a solution with respect to 
these measures exist~\cite{Finkbeiner13,BloemCHJ09,ChatterjeeHJS11,CernyCHRS11
,CernyH11,Zimmermann13}. 
We want to discuss at the SYNT workshop whether any of them should be a 
standard quality measure in future competitions, or whether we should use 
optimization towards them as special challenges for some competitions.

\subsection{Witnesses for Correctness}
\label{sec:witnesses}

The problems that we consider in \syntcomp are realizability of a specification and 
synthesis of a solution. While the production of solutions is optional in 
some other subfields of automated reasoning and computer-aided verification, 
it is at the heart of \syntcomp. Because of this, solutions themselves are 
natural witnesses for the correctness of a ``realizable'' statement. To 
verify that a solution is correct, it can be model checked against the 
specification. 

However, there are a number of problems with this approach:

\begin{enumerate}
\item \label{problem:unreal} solutions of the synthesis problem can only be 
used as witnesses for correctness if the specification is realizable. If it 
is unrealizable, \syntcomp thus far did not require any witness of correctness.
\item \label{problem:mc} model checking may not be easy for complex solutions 
and specifications. Even for safety specifications, we had a number of 
solutions in \syntcomp 2014 that could not be model checked, and an increased 
number in 2015.
\end{enumerate}

In the following, we present some ideas how to handle these problems. Note 
that the proposed solutions are orthogonal and could be combined.

\paragraph{Witnesses for Unrealizability (Counter-strategies):}
To solve problem~\ref{problem:unreal}, the competition could include (either 
by default, or as separate track, or as a challenge) the computation of 
counter-strategies for unrealizable specifications. 

The easiest way to investigate the performance of tools on this task would be 
to run the tool on the negated specification and require a Moore-type 
implementation (instead of the usual Mealy-type). Combining the two 
tasks gives even more meaningful results, as in general we will not know 
whether our specification is realizable or not, and we want a witness of the 
fact regardless of the outcome.

\paragraph{Comprehensive Witnesses for Effective Correctness Proofs:}
Based on our findings in \syntcomp 2014 and 2015, we introduced the possibility that tools (in the AIGER-based safety track) can give 
additional witness information that will make it easier to check correctness 
of the provided solution. For safety specifications, this information can 
simply be an inductive invariant of the produced solution, i.e., a set of 
states that does not contain error states and is such that the produced 
solution will never leave the set. The winning region of the system (computed 
by the standard fixpoint-based algorithm for safety games) is an 
example of an inductive invariant. In \syntcomp 2016, we allow tools to provide such an 
invariant (also in the AIGER format). As mentioned before, this solves the problem of verifying more complex solutions in most cases. 

For specifications in LTL or GR(1), we found the problem of verifiability to 
be even worse. Furthermore, comprehensive witnesses need to contain more 
information if specifications are not restricted to safety. Since the hardest part of the verification of 
liveness properties is essentially the construction of (some form of) 
suitable ranking function, it would be good if this ranking function could 
be supplied by the synthesis tool. In case of GR(1), such ranking 
functions have a rather simple form, which might boil down to 
a fixed unrolling of liveness properties and then effectively checking a 
safety property.

\subsection{Technical Setup}
\label{sec:future-setup}

Both \syntcomp 2015 and 2016 were run at Saarland University, on a small set of machines that were acquired specifically for this purpose. The benefit of this approach is that we were able to tailor the computers to the needs of our competition, which is CPU- and memory-intensive, but does not have a focus on parallelization. For instance, since none of the tools in the competition used more than 3 or 4 cores (in \syntcomp 2014 and 2015, respectively), we had a huge benefit from moving from machines with 16 CPU cores, but low sequential speed (in 2014), to machines with only 4 CPU cores, but nearly twice the sequential speed (in 2015 and 2016). 
Moreover, the organizers have full control over these machines (as opposed to machines that are operated and serviced by a third party), which makes the execution of the competition easier and more predictable. 

However, the reduced number of available competition servers was already an issue last year. To cope with the problem, we reduced the number of benchmark instances that were tested in the competition overall.\footnote{In fact, we significantly reduced the number from 569 to 250 in the realizability track, while increasing the number from 157 to 239 instances in the synthesis track.} In $2016$, the capacity of the competition servers was at its limit, and we had to run experiments until $2$ days before the presentation, instead of having a longer window between execution of presentation as planned. 
Since we want to add another track to the competition next year, a bigger computing capacity is certainly desirable and possibly necessary. 

This could be achieved in different ways, each with their own advantages and disadvantages. Increasing the number of machines in the current setup requires 
dedicated funding and local infrastructure, which may be hard to justify for 
a service that only runs 2-3 months per year. The other option is to use 
third-party machines, for example those provided by the StarExec 
platform~\cite{StumpST14}. The benefit would be essentially unlimited compute 
capacity, while the downside would be that we give up complete control over the 
machines and the execution of the competition, and have to adjust 
our technical setup to the infrastructure of that service (to a degree that is 
currently unknown to us).

%
%
%
%
%
%
%
%

%% file: conclusions.tex
\section{Conclusions}
\label{sec:conclusions}

The Reactive Synthesis Competition has been held annually since 2014.
\syntcomp 2016 presents the biggest extension of the competition thus far, 
introducing an additional track with specifications in full LTL. \syntcomp is designed as a long-term effort that is guided by feedback from the reactive synthesis community, and we will continue to extend and modify the competition to foster the research in scalable and mature implementations of synthesis techniques.

{\small
\myparagraph{Acknowledgments}
We thank R\"udiger Ehlers, Ioannis Filippidis, Ayrat Khalimov, Felix Klein, Andrey Kupriyanov, 
Kim Larsen, Nir Piterman, Markus Rabe, and Leander Tentrup for interesting 
suggestions for the future of \syntcomp (and apologize if we forgot someone).
Finally, we thank Jens Kreber for 
technical assistance during setup and execution of \syntcomp 2016 at 
Saarland University.

The organization of \syntcomp was supported by the Austrian Science Fund
(FWF) through project RiSE (S11406-N23), and by the German
Research Foundation (DFG) through project ``Automatic Synthesis of Distributed and
Parameterized Systems'' (JA 2357/2-1).
}